# PHASE-SPACE CONSTRAINTS ON NEUTRINO LUMINOSITIES

C. SIVARAM, KENATH ARUN[1], SAMARTHA C A

INDIAN INSTITUTE OF ASTROPHYSICS

BANGALORE, 560034

**Abstract**

While the importance of phase space constraints for gravitational clustering of neutrinos (which are fermions) is well recognized, the explicit use of such constraints to limit neutrino emission from ultra energetic sources has not been stressed. Special and general relativistic phase space constraints are shown to limit neutrino luminosities in compact sources in various situations.

---

[1] Christ Junior College

## 1. Introduction

The most energetic events occurring in the Universe such as supernovae and gamma ray bursts not only involve the emission of prodigious amounts of electromagnetic radiation of all bands in a short time duration but are also expected to be accompanied by corresponding vast energy generation in the form of neutrinos.

As is well known, in the late stages of the evolution of a massive star neutrino emission dominates (reducing the lifetime of the star when it starts fusing heavier elements). Processes such as photoneutrino production and pair annihilation into neutrinos dominate at core temperatures of more than a billion degrees.

The core collapse of such a massive star results in a Type II supernova and formation of a neutron star (or black hole) as remnant. The protoneutron star cools from a temperature of a few billion degrees and most of the binding energy of the neutron star $(3 \times 10^{53} \, ergs)$ is radiated in the form of neutrinos in a period of a few seconds.

Again gamma ray bursts releasing $10^{51} - 10^{53}$ ergs in a few seconds are also expected to release an intense burst of neutrinos during the same period. [1]
However unlike photons, neutrinos being fermi particle are constrained by phase space (i.e., two particles per unit volume of phase space, i.e., $\hbar^3$, $\hbar = \dfrac{h}{2\pi}$ being Planck constant).
Many of the discussions of intense neutrino bursts and emissions (e.g. ref. [1-4]) do not seem to consider this aspect.

As we shall see, these constraints limit the neutrino luminosity in various situations. When discussing neutrinos as dark matter (DM) candidates (especially for galactic halos), it is well recognised that taking into account that the neutrinos, being fermions, phase space densities must satisfy the inequality $d^3x.d^3p \geq \hbar^3$, constrains the neutrino mass. For example see [5,6,7].

More precisely for a given dark matter density $\rho_{DM}$ (assumed to be dominated by say neutrinos of mass $m_\nu$, so that $\rho_{DM} = m_\nu n_\nu$) and a typical galactic (constant) rotation velocity of $v$, this constraint on the mass $m_\nu$, translates into,

$$m_\nu \geq \left(\hbar \rho_m / v^3\right)^{1/4}.$$

(For a typical $v \approx 200 km/s; \rho_{DM} \approx 10^{-24} gm/cm^3$, this gives $m_\nu \geq 50 eV$).

Similar phase space constraints on $m_\nu$ arise from neutrino clustering in the gravitational potential of galaxy clusters etc.

While the importance of phase space constraints for gravitational clustering of neutrinos is well recognised [5], the explicit use of such constraints to limit neutrino emission for ultra energetic sources has not been highlighted.

**2. General Case**

Let us consider a general example first. Consider neutrinos of average energy $E_\nu$ being emitted from a source. Let $n_\nu$ number density of the emitted neutrinos produced. Then the phase space constraint $d^3x.d^3p \geq \hbar^3$, translates into:

$$n_\nu^{-1}\left(\frac{E_\nu}{c}\right)^3 \geq \hbar^3.$$

Giving, $n_\nu \leq \left(\dfrac{E_\nu}{\hbar c}\right)^3$ …(1)

Further if $\varepsilon_\nu$ is the energy density of the emitted neutrinos, i.e., $\varepsilon_\nu = n_\nu E_\nu$, then we have:

$$\varepsilon_\nu = \left(\frac{E_\nu}{c}\right)^3 \frac{1}{\hbar^3} E_\nu = \frac{E_\nu^4}{(\hbar c)^3} \qquad …(2)$$

Equation (2) then implies the phase space constraint on the neutrino flux as:

$$f_\nu \leq \frac{E_\nu^4}{4c^2} \frac{1}{\hbar^3} \qquad …(3)$$

For $E_\nu \approx 10 MeV$, this implies,

$$f_\nu \leq 3 \times 10^{39} ergs/cm^2/s \qquad \ldots(4)$$

Notice the $E_\nu^4$, energy dependence!

(For 1 MeV, this is about $f_\nu \leq 3 \times 10^{35} ergs/cm^2/s$).

If the binding energy of the neutron star $\left(\approx 3 \times 10^{53} ergs\right)$ is released by neutrino emission over a period of about ten seconds [7], equation (4) would constraint the source size to be about a few kilometres and also give an upper limit to the source temperature if the emission is thermal.

These constraints are consistent with the intense emission of the neutrinos by a protoneutron star (like in the case of SN1987A).

More generally, if the neutrino diffusion time is $t_d$, then if $\kappa_\nu$ is the neutrino opacity; $n$ is the number density of neutrons, then;

$$t_d \approx \frac{1}{n\sigma_\nu c}; \quad \kappa_\nu \sim nm_p\sigma_\nu \qquad \ldots(5)$$

Equations (3)-(5) can be used to constrain source size. Using a total neutrino opacity [8], $\kappa_\nu = 2 \times 10^{-17} \rho \left(kT_\nu/4MeV\right)^2 cm^{-1}$, we can further refine the constraint.

The outgoing neutrinos could accelerate (by exerting pressure) a spherical shell of matter of radius $r$ and thickness $dr$. This acceleration can be expressed as,

$$a_\nu = \frac{\kappa_\nu}{4\pi r^2 \rho dr} \frac{L_\nu}{c} dr, \quad \rho = nm_n \qquad \ldots(6)$$

$L_\nu$ would be constrained by equations (1)-(4). If the number of neutrino scatterings is $N$, then,

$$L_\nu \approx 4\pi \left(N^{1/2} n^{-1/2} \sigma_\nu^{-1/2}\right)^2 \frac{E_\nu^4}{4c^2} \frac{1}{\hbar^3} \approx \frac{4\pi N}{\sigma_\nu n} \frac{E_\nu^4}{4c^2 \hbar^3} \approx \frac{\pi N E_\nu^4}{\sigma_\nu n c^2 \hbar^3} \qquad \ldots(7)$$

Plugging in the values we get,

$L_\nu \approx 2 \times 10^{52} \, ergs/\sec$.

This can be used in equation (6) to give the value of the acceleration as,

$a_\nu \approx 2 \times 10^6 \, cm/s^2$.

## 3. General Relativistic Case

For compact objects like neutron stars, where General Relativistic effects could be substantial we could use the corresponding generalised phase space relation,

$$\left(\sqrt{g_{00}} d^3 x\right)\left(\sqrt{g_{00}} d^3 p\right) \geq \hbar^3 \qquad \ldots(8)$$

Assume same metric coefficient for both co-ordinate and momentum space.
Where, the coefficients of the Schwarzschild metric are,

$$g_{00} = \left(1 - \frac{2GM}{rc^2}\right); \quad g_{11} = \frac{1}{\left(1 - \frac{2GM}{rc^2}\right)}; \quad g_{22} = r^2; \quad g_{33} = r^2 \sin^2\theta.$$

The phase space constraint in this case gives the limit on the number density as,

$$n_\nu \leq \left(1 - \frac{2GM}{rc^2}\right)\left(\frac{E_\nu}{\hbar c}\right)^3 \qquad \ldots(9)$$

The reduction in the energy due to the redshift is given by,

$$\frac{\Delta E}{E} = \frac{GM}{rc^2}.$$

For a typical $M = 1.4 M_\odot; r = 10 Km$, this shift is of the order of $0.1E$.

The upper limit for the energy density in this case is given by,

$$\varepsilon_\nu \leq \left(1 - \frac{2GM}{rc^2}\right) \frac{E_\nu^4}{(\hbar c)^3},$$

The maximum flux is given by,

$$F \leq \frac{c}{4} \varepsilon_\nu \qquad \ldots(10)$$

For $E_\nu \approx 10 MeV$, this implies,

$$f_\nu \leq 5.5 \times 10^{38} ergs/cm^2/s \qquad \ldots(11)$$

Again considering the binding energy of the neutron star $(\approx 3\times 10^{53} ergs)$ released by neutrino emission over a period of about ten seconds, equation (11) would constraint the source size to be about *30* kilometres.

In certain cases, the entire energy of the burst could be emitted in the form of neutrinos, giving a silent burst in gamma rays, like in the case of collision of neutron stars. The kinetic energy $(\approx 10^{53} ergs)$ of the two NS as well as their binding energy $(\approx 3\times 10^{53} ergs)$ will be released in the burst, giving a total energy released in the form of neutrinos as, $\approx 7\times 10^{53} ergs$.

**4. Special Relativistic Case**

Due to the high velocities of the emitted neutrinos, we have to take into account the Special Relativistic corrections on the phase space constraints, $d^3x.d^3p \geq \hbar^3$.

The deceleration radius of the gamma ray burst gives the co-ordinate space. For a neutrino energy of $E_\nu$ and number density *n*, the co-ordinate space is given by,

$$d^3x = R_D^3 = \left(\frac{3}{4\pi}\frac{E_\nu}{\Gamma^2 nm_P c^2}\right) \text{ and the momentum space } \left(\frac{E_\nu}{\Gamma c}\right).$$

Including these Special Relativistic effects the phase space constrain becomes,

$$\left(\frac{3}{4\pi}\frac{E_\nu}{\Gamma^2 nm_P c^2}\right)\left(\frac{E_\nu}{\Gamma c}\right)^3 \geq \hbar^3.$$

Which gives the limit on the number density of the neutrinos as,

$$n \leq \left( \frac{3}{4\pi} \frac{E_\nu}{\Gamma^2 m_P c^2} \right) \left( \frac{E_\nu}{\hbar^3 \Gamma c} \right)^3 \qquad \ldots(12)$$

And the limit on the energy density as,

$$\varepsilon_\nu \leq \left( \frac{3}{4\pi} \frac{E_\nu}{\Gamma^2 m_P c^2} \right) \left( \frac{E_\nu}{\hbar^3 \Gamma c} \right)^3 E_\nu \qquad \ldots(13)$$

Notice the $E_\nu^5$ dependence and also note the sharp $\Gamma^5$ dependence!

The phase space constraint on the flux from equation (13) is given by,

$$f_\nu \leq \frac{c}{4} \varepsilon_\nu \qquad \ldots(14)$$

For $E_\nu \approx 10 MeV$; $f_\nu \leq 4 \times 10^{36} ergs/cm^2/s$ $\qquad \ldots(15)$

Considering the binding energy of the neutron star $(\approx 3 \times 10^{53} ergs)$ being released by neutrino emission over a period of about ten seconds, equation (15) would constraint the source size to be about *200* kilometres.

Neutrinos being fermions will obey Fermi-Dirac statistics. Hence the maximum allowed energy is the Fermi energy $E_F$.

For a neutrino emission in a cone of angle $\theta_\nu$, the energy flux is given by,

$$F_\nu = \int_0^{\theta_\nu} d\Omega \int_0^{E_F} cE_P^2 \frac{dp}{\hbar^3} = \frac{1-\cos\theta}{16\pi^3} c\varepsilon_\nu.$$

Where, the energy density is given by, $\varepsilon_\nu = E_F \left( p_F / \hbar \right)^3$.

This gives an upper limit on the neutrino luminosity.

$$4\pi R_0^2 F_\nu = R_0^2 \frac{\Gamma^2}{2} c\varepsilon_\nu$$

$$R\Gamma^2 / 2 \leq c\Delta t.$$

The bursts energy is given by,

$$E_\nu < L_\nu \Delta t \frac{\Gamma^2}{2}.$$

If the neutrino burst energy is constrained by the binding energy of the neutron star $(3 \times 10^{53} ergs)$, then the burst duration is given by,

$$\Delta t = \frac{2E_\nu}{\Gamma^2 L_\nu}.$$

Including both special relativistic and general relativistic effects into account, we have the neutrino flux given by,

$$f_\nu \leq \frac{c}{4}\left(1 - \frac{2GM}{rc^2}\right)\left(\frac{3}{4\pi}\frac{E_\nu}{\Gamma^2 m_P c^2}\right)\left(\frac{E_\nu}{\hbar^3 \Gamma c}\right)^3 E_\nu.$$

The comoving luminosity is given by,

$$L_\nu = f_\nu 4\pi R^2.$$

In the observer's frame it is given by,

$$L_\nu = f_\nu 4\pi R^2 \Gamma^2 = \frac{c}{4}\left(1 - \frac{2GM}{rc^2}\right)\left(\frac{3E_\nu}{\Gamma^2 m_P c^2}\right)\left(\frac{E_\nu}{\hbar^3 \Gamma c}\right)^3 E_\nu R^2 \Gamma^2.$$

## 5. Modelling GRB

The phase space constraints obtained for the neutrino flux can be used to model the gamma ray burst, by determining the limit on the energy density.

The neutrinos will undergo pair neutrino annihilation:
$$\nu_e + \bar{\nu}_e \to e^+ + e^- + \gamma + \gamma$$

The minimum energy required for this interaction is given by,
$$2m_e c^2 \approx 1 MeV.$$

The cross section for this interaction is given by,

$$R = \frac{G_F^2 E^2}{(\hbar c)^4} n_e \times n_e c.$$

Where, $R = \sigma v n^2$.

And in equilibrium condition the electron and positron densities are given by,

$$n_{e^+} = n_{e^-} = n_e = \left(\frac{kT}{\hbar c}\right)^3.$$

The energy density is then obtained as,

$$\varepsilon = R(kT)$$

$$\varepsilon = \frac{G_F^2 E^2}{(\hbar c)^4}\left(\frac{kT}{\hbar c}\right)^3 \times \left(\frac{kT}{\hbar c}\right)^3 c(kT) = 10^{-56}\left(\frac{T}{1K}\right)^9 ergs/cm^3.$$

The flux is given by,

$$f = \frac{c}{4}\varepsilon = 7.5 \times 10^{-47}\left(\frac{T}{1K}\right)^9 ergs/cm^2/s.$$

The flux obtained from the phase space constraint is of the order of $4 \times 10^{34} ergs/cm^2/s$.

For the two to be comparable, the temperature should be of the order of $10^9 K$.

One of the mechanisms suggested for the production of intense gamma flux in gamma ray bursts is the neutrino annihilation reaction:

$$\nu + \bar{\nu} \rightarrow \gamma + \gamma,$$

This is especially relevant perhaps for the short duration bursts where merger of two neutron stars is expected to occur. The combined binding energy could be released substantially as neutrino pairs, which could annihilate to produce gamma rays.

It follows that phase space constraints on the neutrino and antineutrino fluxes would in turn imply a constraint on the gamma ray flux arising from this mechanism.

Interactions of hadrons in the region of gamma ray burst (close to central engine and in the jet) could produce high-energy neutrinos and high-energy gamma rays:

$$p + p \rightarrow p + p + \pi^0$$
$$\pi^0 \rightarrow 2\gamma$$
$$n \rightarrow p + e^- + \bar{\nu}_e$$
$$\pi^+ \rightarrow \mu^+ + \nu_\mu$$
$$\mu^+ \rightarrow e^+ + \nu_\mu + \nu_e$$

Phase space constraints apply to the high-energy neutrinos (resulting from hadron decays, etc.) and the maximum neutrino energy released is given by,

$$\frac{G_F^2 E_t^2 E_\nu^4}{\hbar^7 c^6} = \left(\frac{G_F E_t}{\hbar^2 c^2}\right)^2 \frac{c\varepsilon_\nu}{\Gamma^4} \, ergs/\sec.$$

In the following process, $\pi \to \mu + \nu$, the pions produced by proton-proton or proton-photon collisions ($p + p \to p + p + \pi^0$, $p + p \to p + p + \pi^+ + \pi^-$, etc. $p + \gamma \to p + \pi^+ + \pi^-$, etc.) would be of high energy.

For example, for a pion energy of $E_\pi = 200 GeV$, the decay neutrino energy is given by:

$$E_\nu = \left(\frac{m_\pi^2 - m_\mu^2}{m_\pi^2}\right) E_\pi \approx 85 GeV.$$

The maximum limit for the neutrino momentum is given by,

$$P_\nu = \Gamma(1 + \beta) E_\nu / c \approx 2 \times 10^5 \, GeV/c = 2 \times 10^2 \, TeV/c.$$

For $\beta \approx 1; (v \approx c)$ and $\Gamma = 1000$.

This would also apply for other high energy $(TeV)$ neutrino sources like microquasars, blazars, etc. [10] with implications for the expected detectability of fluxes of such particles (at various distances) in large-scale neutrino detectors like Amanda, Icecube, etc. [8] The estimates of constraints and fluxes would be taken up in a later publication [17].

## 6. Modification in the Phase Space Constraint due to Generalised Uncertainty Principle (GUP)

At around the Planck scale, one expects a modification of the uncertainty principle, giving a so called generalised uncertainty principle (GUP).

The 'smooth' structure of space-time manifolds could undergo drastic changes at the Planck scale, perhaps giving rise to discrete structure of space-time, where even spatial co-ordinates become non-commutative, i.e. $[x, y] = iL_{pl}^2$, apart from $[x, p_x] = i\hbar$, etc.

These quantum fluctuations in space-time, could have microscopic consequences which in principle could be detected in deviations from Newton's gravity law, corrections in atomic spectroscopy, with consequences for gravitational wave detectors (in their displacement noise spectrum), k-meson decays, time delays in gamma ray bursts, etc [18,19].

A typical GUP relation (also suggested in superstring theories) is [20]:

$$\Delta x \Delta p \geq \hbar + \frac{\lambda}{\hbar}(\Delta p)^2 \qquad \ldots(16)$$

Where, $\lambda$ has the dimensions of $L_{pl}^2$, where, $L_{pl}$ is the Planck length and,

$L_{pl}^2 \approx 10^{-66} cm^2 \approx 1$ atto shed; 1 shed $= 10^{-24}$ barns

The usual phase space giving the total number of quantum states is modified from $\frac{dV d^3 p}{(2\pi\hbar)^3}(\Delta x \Delta p \sim 2\pi\hbar)$, to, $\frac{dV d^3 p}{(2\pi\hbar)^3(1+\lambda p^2)^3}$, $p^2 = p_i p^i, i = 1,2,3$.

The non-commutativity $[x, p]$ becomes:

$[\hat{x}, \hat{p}] = i\hbar(1+\lambda p^2)$, which can be interpreted as a 'modification' of $\hbar$ to $\hbar' = \hbar(1+\lambda p^2)$ (this was first suggested in ref [21]).

The above relation implies a modification to the phase space constraint given by $d^3x.d^3p \geq \hbar^3$

The GUP (eq.16), implies a phase space constraint of the form,

$$d^3x.d^3p \geq \hbar^3(1+\lambda p^2)^3 \qquad \ldots(17)$$

The number density of the neutrinos will be modified from eq. (1) to the form,

$$n_v^{-1}\left(\frac{E_v}{c}\right)^3 \geq \hbar^3\left(1+\lambda p^2\right)^3 \qquad \ldots(18)$$

Since $\lambda p^2 \ll 1$, we can expand the right hand side of eq. (18) binomially. This gives,

$$n_v^{-1}\left(\frac{E_v}{c}\right)^3 \geq \hbar^3\left(1+3\lambda p^2 + \ldots\right) \text{ (Neglecting higher order terms in } \lambda p^2 \text{)}.$$

This gives the upper limit on number density as,

$$n_v \leq \left(\frac{E_v}{\hbar c}\right)^3 \left(1+3\lambda p^2\right)^{-1} \qquad \ldots(19)$$

The phase space constraint on the neutrino flux will be modified from that of eq. (3) to,

$$f_v \leq \frac{E_v^4}{4c^2 \hbar^3}\left(1 - \frac{3\lambda}{\hbar^2}\left(\frac{E_v}{c}\right)^2\right) \qquad \ldots(20)$$

Where, the neutrino momentum is given by, $p = \dfrac{E_v}{c}$

The first term corresponds to the phase space constraint given by eq. (3), and the second term corresponds to the modification introduced by the GUP.

For neutrino energies of the order of $TeV$, and $\lambda \approx L_{pl}^2 \approx 10^{-66} cm^2$, the modification is of the order of $10^{-32}$.

The highest observed energies (in cosmic rays) are of the order of $10^9 TeV$. The correction corresponding to this is of the order of $10^{-14}$.

As we can see, the effects are too small to be detected.

If the scale is reduced to the order of weak interactions, that is, $\lambda \approx l_W^2 \approx 10^{-33} cm^2$, for $TeV$ energies, the correction is of the order of 10. Since this modification is very large, we can conclude that the assumed reduction in the scale does not take place.